\documentclass[12pt]{article}
\usepackage[reqno]{amsmath}
\usepackage{epsfig}
\usepackage{array}
\usepackage{float}
\usepackage{lscape,graphicx}
\usepackage{graphics}
\usepackage{amssymb}
\usepackage{color}

\usepackage{array}
\newcolumntype{L}[1]{>{\raggedright\let\newline\\\arraybackslash\hspace{0pt}}p{#1}}
\newcolumntype{C}[1]{>{\centering\let\newline\\\arraybackslash\hspace{0pt}}p{#1}}
\newcolumntype{R}[1]{>{\raggedleft\let\newline\\\arraybackslash\hspace{0pt}}p{#1}}

\newcommand{\bea}{\begin{eqnarray}}
\newcommand{\eea}{\end{eqnarray}}
\newcommand{\ea}{\end{array}}

\newcommand{\be}{\begin{equation}}
\newcommand{\ee}{\end{equation}}
\newcommand{\bad}{\begin{array}{ccc}}

\newcommand{\ba}{\begin{array}{c}}
\begin{document}

\bibliographystyle{ieeetr}

\begin{center}
\mathversion{bold}
{\bf{\large The $\mu\to e\gamma$ decay in an EW-scale non-sterile RH neutrino model}}
\mathversion{normal}

\vspace{0.4cm}
D. N. Dinh\\
%
\vspace{0.1cm}
{\em Institute of Physics, Vietnam Academy
of Science and Technology, \\
10 Dao Tan, Hanoi, Vietnam.\\
}
%

\end{center}
\begin{abstract}
\noindent We study in this research the phenomenology of $\mu\to e\gamma$ decay in a scenario of the class of extended models with non-sterile right-handed (RH) neutrino at electroweak (EW) scale proposed by P.Q. Hung. Field content of the standard model (SM) is enlarged by introducing for each SM fermion a corresponding mirror partner with the same quantum numbers beside opposite chirality. Light neutrino masses are generated via the type-I see-saw mechanism and it is also proved to be relevant with low energy within the EW scale of the RH neutrino masses. We introduce the model and derive branching ratio of the $\mu\to e\gamma$ decay at one-loop approximation with the participation of W gauge boson, neutral and singly charged Higgs scalars. After that we set constraints on relevant parameters and predict the sensitivities of the decay channel under the present and future experiments. 
\end{abstract}

\section{Introduction}
	
	The electroweak-scale right-handed neutrino (EW-scale $\nu_R$) model is an extended version of the standard model (SM) was proposed for the first time in \cite{Hung:2006ap}. The fermion contents are doubled by introducing a mirror partner for each of the SM particle. Thus, correspond to a normal fermionic component is a mirror one with opposite chirality. Obey the mirror symmetry, right-handed mirror neutrinos and leptons, for example, are combined to form a doublets of the model's gauge symmetry $ SU(2)\times U(1)_Y$. Other particles in the mirror sector are arranged in the similar way.\\

\noindent With left-handed and right-handed neutrinos respectively introduced in the normal and mirror sectors, that acquires enough conditions for the type I see-saw mechanism operating to give masses for the light active neutrinos \cite{Minkowski:1977sc, Gell-Mann:1979vob, Yanagida:1979as, Mohapatra:1979ia}. In contrast to the normal type I see-saw, in which Dirac mass matrix is generated at electroweak scale, therefore right-handed neutrino masses, in general, should be extremely heavy to ensure the SM active neutrino masses as small as experiments have identified, Dirac mass matrix in the current model is given by a new Higgs singlet apart from the mechanism of SM mass production. It is proved that if the Higgs singlet's VEV is at relevant scale, heavy Majorana mass matrix is about hundred GeV, thus at the EW scale \cite{Hung:2006ap}.\\

\noindent Apparently, mirror partners of the SM matter particles introduced in 
this model, which might impact on various physical phenomena, have to confront 
with experimental high precise measurements of the EW processes. The effects of extra 
chiral doublets have been carefully examined in \cite{Hoang:2013jfa}. The research 
shows that there is still large free parameter space after being constrained by the 
EW precision data. One of the exciting reasons, which has been demonstrated, is the partial cancellation of the contributions from the mirror fermions by those of the physical scalars, especially the $SU(2)$ triplets. An updated version has been 
introduced after the discovery of the $125$ GeV SM-like scalar \cite{Aad:2012tfa, Chatrchyan:2012ufa}, which has opened up a new stage for elementary particle physics,  particularly model building for physics beyond the SM. Differ from the old version, 
an additional Higgs doublet has been introduced to give masses for mirror quarks and charged leptons (mirror sector), and the original one for those of the normal sector.
Two candidates are found out to have signals in agreement with ATLAS and CMS observations  \cite{Hoang:2014pda}.\\

\noindent Inspire of compelling evidence for lepton flavour violations (LFV) in 
neutrino oscillations, all efforts looking for those in the charged lepton sector 
have given negative results so far. In fact, it is demonstrated that minimal extensions of the SM with massive neutrinos, the rates of LFV processes involving the charged leptons are so extremely tiny, that unobservable in practice. For instance, the $\mu\to e\gamma$ decay branching ratio is evaluated to be about $10^{-55}$ using the currently known neutrino oscillation data \cite{Petcov:1976ff, Bilenky:1977du, Cheng:1980tp}. For the reason, the decay is considered as one of the most important channels to look for 
signals of physics beyond the SM. On the experimental aspect, the best upper limit implies from the non-observation of the muon decay $\mu\to e\gamma$ given by MEG in 
2016
\begin{equation}
\label{mutoegexp}
{\rm BR}(\mu^+\rightarrow e^+\gamma)
< 4.2\times 10^{-13}~~\cite{TheMEG:2016wtm}\,,\\
\end{equation}
and  it has been recently upgraded to work at sensitivity
\begin{equation}
\label{mutoegexpUpg}
{\rm BR}(\mu^+\rightarrow e^+\gamma)
< 6.0\times 10^{-14}~~\cite{MEGII:2021fah,Renga:2014xra}.\\
\end{equation}
On the theoretical aspect, a large number of researches have been interested 
in the same topic in various scenarios of physics beyond the SM \cite{Petcov:1976ff, Bilenky:1977du, Cheng:1980tp, Hisano:1995cp, Alonso:2012ji, Kakizaki:2003jk, Akeroyd:2009nu, Leontaris:1985qc, Raidal:1997hq, Ma:2000xh, Petcov:1982en, Abada:2007ux, Hung:2007ez, Hung:2015hra, Chang:2017vzi, Dinh:2012bp, Dinh:2019jdg, Huong:2019vej}. Among them, the $\mu\to e\gamma$ branching ratios and related consequences in the considering schemes of EW-scale $\nu_R$ models are discussed for three different versions, which are the original \cite{Hung:2007ez}, the updated with light of the $125$ GeV SM-like Higgs scalar discovery \cite{Hung:2015hra} and an extension with $A_4$ discrete symmetry \cite{ Chang:2017vzi}, respectively.\\

\noindent Although the phenomenology of $\mu\to e\gamma$ decay in the three mentioned above EW-scale $\nu_R$ scenarios has been considered, these researches have only taken into account contributions of one-loop diagrams, which are formed by light neutral scalar and the mirror charged leptons. The theoretically predicted branching ratio  confronted with experimental upper bound will set upper limits on the magnitudes of Yukawa couplings involving the Higgs singlet. Following that light singlet vacuum expectation value $v_s$ is identified to have right Dirac mass magnitude for the see-saw mechanism working properly with hundred GeV Majorana right-handed neutrino masses to generate sub-eV scale of those for the light active neutrinos. However it is only part of the whole story, at one-loop approximation there are still contributions from diagrams with $W^-$ gauge boson, other neutral and singly charged scalars. These contributions might be sizable and therefore required careful considerations to set constraints on the involving interaction strengths, or evaluate their observed possibilities with the current and future experiments.\\

\noindent As aim of this research, we discuss the phenomenology of $\mu\to e\gamma$ decay in the scenario of EW-scale $\nu_R$ model with two Higgs doublets, the extended version to accommodate with the $125$ GeV SM-like scalar detection \cite{Hoang:2014pda, Hung:2015hra}. The process is considered upto one-loop approximation with participation of light physical scalar and other particles, which have not been studied in the previous 
researches. The paper is divided into 4 main sections as follows. Beside this section 
for introduction, in Sect. 2, we briefly introduce the model and write down the relevant LFV vertexes, which contribute to our process of interest. In Sect. 3, after introducing the form-factors with explicitly algebraic expressions, one will derive the decay branching ratio and perform numerical analysis. Conclusion is given in the Sect. 4.  
 
\section{\label{model} A review of the model}
	
	\subsection{The model content}
	That has been mentioned in an earlier part, this research works on an extended 
	version of the EW-scale $\nu_R$ model, which is constructed based on the symmetric group 
	$SU(2)\times U(1)_Y\times U(1)_{SM} \times U(1)_{MF}$. The $SU(2)\times U(1)_Y$ is the gauge group of the model, 
	and $U(1)_{SM} \times U(1)_{MF}$  is a global symmetry introduced to forbid some unexpected interactions. Matter fields are arranged as following under 
	the gauge group: 
	\begin{center}
	SM particles ~~~~~~~~~~~~~~~~~~~~~~~~~Mirror particles \\
    $\ell_L=(\nu_L,~e_L)^T$,~$e_R$  ~~~~~~~~~~~~~~~~$\ell_R^M=(\nu_R,~e_R^M)^T$,~$e_L^M$ \\
    $q_L=(u_L,d_L)^T$,~$u_R$, $d_R$  ~~~~~~~~~~~~$q_R^M=(u_R^M,~d_R^M)^T$,~$u_L^M$, $d_L^M$.
    \end{center}
    With this choice, in contrast to the SM, right-handed neutrinos are components of a $SU(2)\times U(1)_Y$ doublets, therefore they are non-sterile and take part in the weak interaction.\\
    
    \noindent Two SM-like Higgs doublets are introduced to give masses to the charged and mirror fermions, 
    respectively. The Higgs doublet, which is denoted as $\Phi_2=(\phi_2^+,\phi_2^0)$, couples to the SM 
    fermions to produce their masses, while the another, called $\Phi_{2M}=(\phi_{2M}^+,   \phi_{2M}^0)$, 
    is responsible for mass generations of the matter partner particles in the mirror sector. The mechanisms 
    of lepton mass generations, including both charged leptons and neutrinos, will be detail presented in a 
    later part of the research.\\
    
    \noindent The heavy right-handed neutrinos, in this model, are introduced naturally in the mirror 
    sector, therefore light active neutrino masses are commonly expected to be generated by the type I 
    seesaw mechanism. The right-hand neutrino mass matrix is provided by a complex Higgs triplet
     with $Y=2$
	  \begin{equation}
	  \tilde{\chi}=\frac{1}{\sqrt{2}}\vec{\tau}.\vec{\chi}=
	  \left(\begin{array}{cc}
      \frac{1}{\sqrt{2}}\chi^+ & \chi^{++} \\
      \chi^0 & -\frac{1}{\sqrt{2}}\chi^+ 
       \end{array}\right).
	  \end{equation}     
      If only one triplet, e.g. $\tilde{\chi}$, is introduced in the model, the tree-level 
      result $\rho=1$, which is precisely measured by experiment, will be spoiled out, then one might 
      obtain $\rho=2$, instead. However, it is also proven in \cite{Chanowitz:1985ug} that, if one has two triplets with
      relevant hyper-charges, when combined, to form $(3,3)$ representation under the global 
      $SU(2)_L\otimes SU(2)_R$ symmetry, the custodial $SU(2)$ symmetry is preserved and $\rho=1$. 
      Thus accompanying with the triplet $\tilde{\chi}$, it is required to add a real Higgs triplet with 
      $Y=0$, called $\left(\xi^+,\xi^0,\xi^{-}\right)$. Finally, we introduce SM singlet Higgs $\phi_S$, 
      which is responsible for creating Dirac mass matrix and connecting between matter and mirror 
      sectors.\\ 
    
	 \noindent In this considered model, two Higgs doublets $\Phi_2$ and $\Phi_{2M}$ are used to couple to SM and 
	 mirror fermions, respectively. To prevent unexpected couplings, a global symmetries $U(1)_{SM}\times U(1)_{MF}$ 
	 are imposed. Transformations of matter and Higgs fields are defined as following:\\
	~i)~$U(1)_{SM}:$ $\Psi=\left\{\Phi_2,~q_L^{SM},~\ell_L^{SM}\right\}$ transforms as 
	$\Psi\rightarrow e^{i\alpha_{SM}}\Psi$,\\
	~ii)~$U(1)_{MF}:$ $\Psi=\left\{\Phi_{2M},~q_R^{M},~\ell_R^{M}\right\}$ transforms as 
	$\Psi\rightarrow e^{i\alpha_{MF}}\Psi$,\\
	~iii)~$\phi_S\rightarrow e^{-i(\alpha_{MF}-\alpha_{SM})}\phi_S$,~$\tilde{\chi}\rightarrow e^{-2i\alpha_{MF}}
	\tilde{\chi}$, and the unmentioned fields are singlet under this global symmetry.\\
	
	\noindent Before writing down the Yukawa couplings, we should keep in mind that the global symmetry defined above only allows $\Phi_2$ to couple to SM 
	fermions, while $\Phi_{2M}$ will couple to the mirror ones. Apparently, terms involved $\phi_S$ should 
	contain both SM and mirror fermion fields. Two unit lepton number violation term, which stands in need 
	for acquiring Majorna mass for the operation of the see-saw type I, can only be constructed without 
	the absence of $\tilde{\chi}$. Detail expressions of the Yukawa couplings are:
	\begin{equation}
	\label{YukawaLepton}
	 \mathcal{L}_Y^{\ell}=g_{\ell}\bar{\ell}_L\Phi_2 e_R+g_{\ell}^M\bar{\ell}_R^M\Phi_{2M} e_L^M
	 +g_{\ell s}\bar{\ell}_L\phi_s \ell_R^M + h.c.,
    \end{equation}	
    \begin{equation}
	 \mathcal{L}_Y^{q}=g_u\bar{q}_L\tilde{\Phi}_2 u_R+g_d\bar{q}_L\Phi_2 d_R
	 +g_{u}^M\bar{q}_R^M\tilde{\Phi}_{2M} u_L^M +g_{d}^M\bar{q}_R^M\Phi_{2M} d_L^M
	 +g_{qs}\bar{q}_L\phi_S q_R^M + h.c.,
    \end{equation}	
\begin{equation}
 \mathcal{L}_{\nu_R}=g_{M} l_R^{M,T} \,\sigma_2\, \tilde{\chi} \, l_R^{M} \, ,
\end{equation}    
where $\sigma_2$ is the second Pauli matrix, $\tilde{\Phi}_2=i\sigma_2\Phi_2^*$ and $\tilde{\Phi}_{2M}=i\sigma_2\Phi_{2M}^*$. 
\subsection{Symmetry breaking and mass generations}
We will discuss next the mechanism of mass generations for matter particles in this model, especially for 
leptons including both charged leptons and neutrinos which are involved in the later discussions of the research, 
when the symmetry is spontaneously breaking. For further discussions, we suppose that Higgs fields develop their vacuum expectation values 
(VEV) as following: 
$\langle\Phi_2\rangle=(0,v_2/\sqrt{2})^T$, $\langle\Phi_{2M}\rangle=(0,v_{2M}/\sqrt{2})^T$, 
$\langle\chi^0\rangle=v_M$, and $\langle\phi_S\rangle=v_S$.\\

\noindent Charged lepton mass matrix obtained from eq. (\ref{YukawaLepton}) can be expressed as
\begin{equation}
	\label{LeptonMass}
	M_\ell=\left( \begin{array}{cc}
	m_\ell & m_\ell^D \\
	(m_\ell^D)^\dagger & m_{\ell M}
	\end{array} \right) \,,
\end{equation}
where $m_\nu^D=m_\ell^D=g_{\ell s}v_S$, $m_\ell=g_{\ell} v_2/\sqrt{2}$, and 
$m_{\ell M}=g_{\ell}^M v_{2M}/\sqrt{2}$. The matrix shown in eq. (\ref{LeptonMass}) can be diagonalized to give eigenvalues in the mass basis and mixing matrix for charged lepton. Without losing physical reality and is easier for calculation to obtain algebraic expression of the mixing matrix, let us assume that 
$m_{\ell M}\gg m_\ell$ and $m_{\ell M},m_\ell\gg m_\ell^D$. The assumption allows us to approximately
block diagonalized $M_\ell$ in the same way usually done for the see-saw type I neutrino mass matrix, then one has 
\begin{equation}
\label{BlockCLT}
	M_\ell=\left( \begin{array}{cc}
	m_\ell & m_\ell^D \\
	(m_\ell^D)^\dagger & m_{\ell M}
	\end{array} \right) \,= \left( \begin{array}{cc}
	I & R_\ell \\
	-R_\ell^\dagger & I
	\end{array} \right) \,\left( \begin{array}{cc}
	\tilde{m}_\ell & 0 \\
	0 & \tilde{m}_{\ell M}
	\end{array} \right) \,\left( \begin{array}{cc}
	I & R_\ell \\
	-R_\ell^\dagger & I
	\end{array} \right)^\dagger \,,
\end{equation}
where $R_\ell\approx \frac{m_\ell^D}{ m_{\ell M}}\ll 1$, and
\begin{eqnarray}
&&\tilde{m}_\ell=m_\ell-\frac{(m_\ell^D)^2}{m_{\ell M}-m_{\ell}}\approx m_\ell,\\
&&\tilde{m}_{\ell M}=m_{\ell M}+\frac{(m_{\ell M}^D)^2}{m_{\ell M}-m_{\ell}}\approx m_{\ell M}.
\end{eqnarray}
Suppose that normal and mirror charged lepton matrices are written in form $\tilde{m}_\ell=U_{\ell L}
m_\ell^d U_{\ell R}^\dagger$, $\tilde{m}_{\ell M}=U_{\ell L}^Mm_{\ell M}^d {U_{\ell R}^M}^\dagger$,
where $m_{\ell}^d$ and $m_{\ell M}^d$ are diagonal, we have the relation between the gauge states and 
physical states as following  
\begin{equation}
	\left( \begin{array}{c}
	\ell_{L(R)} \\
	\ell^M_{L(R)}
	\end{array} \right) \,=\left( \begin{array}{cc}
	U_{\ell L(R)} & -R_\ell U_{\ell L(R)}^M \\
	R_\ell^\dagger U_{\ell L(R)} & U_{\ell L(R)}^M
	\end{array} \right) \,\left( \begin{array}{c}
	\ell'_{L(R)} \\
	{\ell^M}'_{L(R)}
	\end{array} \right).
	\label{LepBasisConver}
\end{equation}
Similarly, we have the quark mass matrix 
\begin{equation}
	\label{Mell}
	M_q=\left( \begin{array}{cc}
	m_q & m_q^D \\
	(m_q^D)^\dagger & m_{q M}
	\end{array} \right) \,,
\end{equation}
noticing that $m_q^D=g_{q s}v_S$, $m_q=g_{q} v_2/\sqrt{2}$, and $m_{q M}=g_{q}^M v_{2M}/\sqrt{2}$.
Block diagonalization gives us  
\begin{eqnarray}
&&\tilde{m}_q=m_q-\frac{(m_q^D)^2}{m_{q M}-m_{q}},\\
&&\tilde{m}_{q M}=m_{q M}+\frac{(m_{q M}^D)^2}{m_{q M}-m_{q}}.
\end{eqnarray}
Discussion about quark sector will not be carried on any further, because they do not involve in the LFV decays being considered in this research.\\ 
  
\noindent Denote the heavy Majorana mass matrix as $M_R=g_M v_M$, one easily obtains the full neutrino mass matrix, which has canonical form of the type-I see-saw mechanism  
\begin{equation}
	\label{Mell}
	M_\nu=\left( \begin{array}{cc}
	0 & m_\nu^D \\
	(m_\nu^D)^T & M_{R}
	\end{array} \right) \,.
\end{equation}
Block diagonalizing the matrix, one could rewrite it as the following, for 
$M_{R}\gg m_\nu^D$, 
\begin{equation}
\label{BlockNeMass}
	M_\nu=\left( \begin{array}{cc}
	0 & m_\nu^D \\
	(m_\nu^D)^T & M_{R}
	\end{array} \right) \,= \left( \begin{array}{cc}
	I & R_\nu \\
	-R_\nu^\dagger & I
	\end{array} \right) \,\left( \begin{array}{cc}
	\tilde{m}_\nu & 0 \\
	0 & \tilde{m}_{\nu R}
	\end{array} \right) \,\left( \begin{array}{cc}
	I & R_\nu \\
	-R_\nu^\dagger & I
	\end{array} \right)^T \,,
\end{equation}
where $R_\nu\approx \frac{m_\nu^D}{ M_{ R}}$, and
\begin{equation}
\tilde{m}_\nu\approx -\frac{(m_\nu^D)^2}{M_{R}}=-\frac{(g_{\ell s} v_S)^2}{g_M v_M},~~ 
\tilde{m}_{\nu R}\approx M_{R}.
\end{equation}
Differ from eq. (\ref{BlockCLT}), where charged lepton mass matrix $M_\ell=M_\ell^\dagger$ is Hermitic thus being transformed by an unitary and its conjugated matrices, the complex symmetric  
$M_\nu=M_\nu^T$ is rotated by an unitary and its orthogonal ones (see eq.(\ref{BlockNeMass}) ).
Light neutrino mass matrix $\tilde{m}_\nu$ is experimentally constrained to be smaller than $1~eV$, 
if $(g^2_{\ell s}/g_M)\sim O(1)$ and $v_S\sim O(10^5~eV)$, Majorana mass of the right-handed neutrino 
$M_R$ could be at order of the electroweak scale. In fact, $g_{\ell s}$ is constrained by some
rare processes (for instant $\mu\to e\gamma$ decay, as we will see latter), which might lead to constraint 
$(g^2_{\ell s}/g_M)\ll 1$. In this case $v_S$ would require to be at order GeV to ensure neutrino mass generation 
operating at the electroweak scale.  
\\
Suppose that light and heavy (mirror) neutrino mass matrices are written respectively as 
$\tilde{m}_\nu=U_{\nu}^* m_\nu^d U_{\nu}^\dagger$, $\tilde{m}_{\nu R}={U_{\nu}^M}^*m_{\nu M}^d 
{U_{\nu}^M}^\dagger$, where $ m_\nu^d$ and $m_{\nu M}^d$ are diagonal, we easily obtain
\begin{equation}
	\left( \begin{array}{c}
	\nu_{L} \\
	(\nu_{R})^c
	\end{array} \right) \,=\left( \begin{array}{cc}
	U_{\nu} & -R_\nu U_{\nu}^M \\
	R_\nu^\dagger U_{\nu} & U_{\nu}^M
	\end{array} \right) \,\left( \begin{array}{c}
	\chi_{\nu} \\
	\chi_M
	\end{array} \right).
	\label{NuBasisConver}
\end{equation}
To guarantee $\rho=1$, as mentioned before, the Higgs potential should have a global symmetry 
$SU(2)_L\otimes SU(2)_R$, and it is broken down to the custodial $SU(2)$ when Higgs fields gain their 
VEVs. Before the symmetry broken, two triplets combine to form $(3,3)$ representation and the doublets 
also maintain the $(2,2)$ structures under the global symmetry. The detail expressions are: 
\begin{equation}
	\label{chi}
	\chi = \left( \begin{array}{ccc}
	\chi^{0} &\xi^{+}& \chi^{++} \\
	\chi^{-} &\xi^{0}&\chi^{+} \\
	\chi^{--}&\xi^{-}& \chi^{0*}
	\end{array} \right) \,,
\end{equation}
\begin{equation}
\Phi_2= \left( \begin{array}{cc}
	\phi_2^{0,*} & \phi_2^+ \\
	\phi_2^- & \phi_2^{0}
	\end{array} \right),~~~
	\Phi_{2M}= \left( \begin{array}{cc}
	\phi_{2M}^{0,*} & \phi_{2M}^+ \\
	\phi_{2M}^- & \phi_{2M}^{0}
	\end{array} \right).
\end{equation}
From the above equations, one have the proper vacuum alignment for breaking gauge symmetry 
from $SU(2)_L \times U(1)_Y$ to $U(1)_{em}$
\begin{equation}
\label{chivev}
\langle \chi \rangle = \left( \begin{array}{ccc}
v_M &0&0 \\
0&v_M&0 \\
0&0&v_M
\end{array} \right) \,,
\end{equation}
\begin{equation}
\langle\Phi_2\rangle= \left( \begin{array}{cc}
	v_2/\sqrt{2} & 0 \\
	0 & v_2/\sqrt{2}
	\end{array} \right),~~~
	\langle\Phi_{2M}\rangle= \left( \begin{array}{cc}
	v_{2M}/\sqrt{2} & 0 \\
	0 & v_{2M}/\sqrt{2}
	\end{array} \right).
\end{equation}
Thus, the VEVs of real components of $\Phi_2$, $\Phi_{2M}$ and $\chi$ are $(v_2/\sqrt{2})$, 
$(v_{2M}/\sqrt{2})$ and $v_M$ respectively, they satisfy the condition
\be
	v_2^2 + v_{2M}^2 + 8\,v_M^2 = v^2\,,
\ee
where $v \approx 246 ~GeV$. For further discussion, the following definitions are used 
\be\label{eq:sdefs}
	s_2 = \dfrac{v_2}{v}; ~~s_{2M} = \dfrac{v_{2M}}{v}; ~~s_M = \dfrac{2 \sqrt{2}\; v_M}{v}\,.
\ee
After gauge symmetry is spontaneously broken, the L-R global symmetry of the Higgs potential is also
broken down to the custodial $SU(2)_D$. Seventeen degrees of freedom of the two Higgs triplets 
(one real and one complex) and two Higgs doublets are rearranged into physical Higgs bosons, beside 
three of the Nambu-Goldstone bosons are absorbed to give masses for W's and Z. Among the physical bosons, 
those which have degenerate masses, are grouped in the same physical scalar multiplets of the global custodial 
symmetry as following:
    \begin{eqnarray}
		\text{five-plet (quintet)} &\rightarrow& H_5^{\pm\pm},\; H_5^\pm,\; H_5^0;\nonumber\\[0.5em]
		\text{triplet} &\rightarrow& H_{3}^\pm,\; H_{3}^0;\nonumber\\[0.5em]
		\text{triplet} &\rightarrow& H_{3M}^\pm,\; H_{3M}^0;\nonumber\\[0.5em]
		\text{three singlets} &\rightarrow& H_1^0,\; H_{1M}^0,\; H_1^{0\prime}\,.
	\end{eqnarray}
The above discussion about physical scalars does not depends on the specific case of Higgs potential, but works
for any one of that which possesses $SU(2)_L\otimes SU(2)_R$ global symmetry, including the cases have been detail considered in 
\cite{Hoang:2014pda}. The physical Higgs bosons, which are generated after the process of gauge symmetry breaking,
should have masses at electroweak scale, thus in range of hundred to few hundred GeVs. The scalars arranged in 
the same multiplets (not singlet) have same masses, while three singlets $H_1^0,\; H_{1M}^0,\; H_1^{0\prime}\,$ 
are not physical states, in general. These states are linear combinations of mass eigenstates, which are denoted
respectively as $\tilde{H}_1^0,\; \tilde{H}_2^0,\; \tilde{H}_3^0\,$. Relations between physical and gauge states are 
expressed as $H_1^0=\sum_{i}^3\alpha_i\tilde{H}_i$, $H_{1M}^0=\sum_{i}^3\alpha_i^M\tilde{H}_i$, where 
$\sum_{i}^3|\alpha_i|^2=1$ and $\sum_{i}^3|\alpha_i^M|^2=1$. Note that the SM Higgs 
scalar discovered by LHC with mass 125-GeV is one of the three mentioned above mass states.\\

\noindent Final physical scalar to be mentioned in this research is the light singlet $\phi_s^0$, which 
originates from the degree of freedom of gauge singlet Higss $\phi_S$. As a singlet, $\phi_S$ does not break the gauge symmetry and its VEV $v_S$ is expected to be much lower than the electroweak scale 
in order to give tiny masses for the light active neutrinos. It is reasonable to take $\phi_s^0$ mass at the same order as $v_S$.
	
\subsection{The LFV vertexes}
It is known that LFV vertex does not present in the SM at tree-level, because charged lepton mass matrix and the matrix of Yukawa 
couplings are diagonal at the same time, and vector gauge bosons interact only with the left-handed components of the matter fields. In this considered scenario, vector fields interact not only with the left-handed SM fermions 
but also with the right-handed components of the mirror sector. Moreover, LFV interactions occur at tree-level for both
the charged currents and Yukawa couplings. Lagrangian involving charged currents in this model can be written as

\be
\label{lagrangian1}
{\cal L}^{CC}= {\cal L}^{CC}_{SM} + {\cal L}^{CC}_{M} \,,
\ee
\bea
\label{lagrangian2}
{\cal L}^{CC}_{SM}= -(\frac{g}{2\,\sqrt{2}})\sum_{i}\,\bar{\psi}^{SM}_{i}\,\gamma^{\mu}\,
(1-\gamma_{5})[\tau^{-}\,W^{+}_{\mu} + \tau^{+}\,W^{-}_{\mu}]\,\psi^{SM}_{i} \,,
\eea
\bea
\label{lagrangian3}
{\cal L}^{CC}_{M}=-(\frac{g}{2\,\sqrt{2}})\sum_{i}\,\bar{\psi}^{M}_{i}\,\gamma^{\mu}\,
(1+\gamma_{5})[\tau^{-}\,W^{+}_{\mu}+ \tau^{+}\,W^{-}_{\mu}]\,\psi^{M}_{i} \,,
\eea
where ${\psi}^{SM}$ and ${\psi}^{M}$ stand for the SM and mirror fermionic fields in the gauge basis, respectively.\\ 

\noindent The relevant Yukawa couplings between the leptons and mirror lepton with scalars, which contribute to 
the phenomenology of $\mu\to e\gamma$ decay, in the gauge basis are listed in the 
table \ref{TableVertex0}.
\begin{table}[h]
\centering
	\scalebox{1}{
		\begin{tabular}{|c|c|c|c|}
			\hline
			Vertices & Couplings & Vertices & Couplings\\
			\hline
			$\bar{e} e H^0_1$ & $  -i\frac{m_\ell g}{2M_W s_2}$ 
			& $\bar{e}^M e^M H^0_{1M}$ & $   -i\frac{m_\ell^M g}{2M_W s_{2M}}$ 
			 \\
			\hline
			$\bar{e} e H_3^0$ & $   -i\frac{m_\ell~g~s_M}{2M_W c_M}\gamma_5$ 
			& $\bar{e}^M e^M H^0_3$ & $   i\frac{m_\ell^M~g~s_M}{2M_W c_{M}}\gamma_5$ 
			 \\
			\hline
			$\bar{e} e H_{3M}^0$ & $   i\frac{m_\ell~g~s_{2M}}{2M_W s_2}\gamma_5$ 
			& $\bar{e}^M e^M H^0_{3M}$ & $   -i\frac{m_\ell^M~g~s_2}{2M_W s_{2M}}\gamma_5$ 
			 \\
			\hline
			$\bar{e}_R \nu_L H_3^-$ & $   -i\frac{m_\ell~g~s_M}{\sqrt{2}M_W c_M}$ 
			& $\bar{e}^M_L \nu_R H^-_3$ & $   -i\frac{m_\ell^M~g~s_M}{\sqrt{2}M_W c_{M}}$ 
			 \\
		   \hline
		   $\bar{e}_R \nu_L H_{3M}^-$ & $   -i\frac{m_\ell~g~s_{2M}}{\sqrt{2}M_W s_2 c_M}$ 
			& $\bar{e}^M_L \nu_R H^-_{3M}$ & $   -i\frac{m_\ell^M~g~s_M}{\sqrt{2}M_W s_{2M} c_{M}}$ 
			\\
		   \hline
	\end{tabular}}
	\caption{Vertexes that contribute to the $\ell\to \ell'\gamma$ decay rates, written in the gauge basis.}
	\label{TableVertex0}
\end{table}
Have good consistency with the current experimental observation and for simplicity, we suppose that 
charged lepton and mirror charged lepton mixing matrices are real (thus all involved complex phase are ignored) 
and $U_{\ell L}=U_{\ell R}=U_{\ell}$, $U_{\ell L}^M=U_{\ell R}^M=U_{\ell}^M$.  Using the relations 
described in eqs. (\ref{LepBasisConver}), (\ref{NuBasisConver}), one easily obtains the main vertexes that contribute to our process of interest in the mass eigenstate basis. After ignoring terms, which are proportional to the second order of $R_{\nu(\ell)}$, the detail expressions are listed in table \ref{TableVertex} and eqs. from (\ref{YuV01}) to (\ref{YuV06}):
\begin{table}[h]
\centering
	\scalebox{1}{
		\begin{tabular}{|c|c|}
			\hline
			Vertices & Couplings\\
			\hline
			$(\bar{e}'_L\gamma^\mu \chi_L)W_\mu^-$ & $ -i\frac{g}{\sqrt{2} }U_{W_\mu}^{L}= -i\frac{g}{\sqrt{2} }U_{PMNS}$
			\\
			\hline 
			$(\bar{e}'_L\gamma^\mu \chi_L^M)W_\mu^-$ & $  i\frac{g}{\sqrt{2}}U_{W_\mu}^{ML}= i\frac{g}{\sqrt{2} }
			\tilde{R}_\nu \left(U_{PMNS}^{M}\right)^*$ 
			 \\
			\hline
			$(\bar{e}'_R\gamma^\mu \chi_L^c)W_\mu^-$ & $  -i\frac{g}{\sqrt{2}}U_{W_\mu}^{R}= -i\frac{g}{\sqrt{2} }
			\tilde{R}_\nu^T \left(U_{PMNS}\right)^*$
			 \\
			\hline
			$\bar{e}'_R\chi_L  H_3^-$ & $  -i\frac{g}{2}Y_{H_3^-}^{L}= -i\frac{g~s_M}{2M_W c_M}m_{\ell}^dU_{PMNS}$ 
			\\
			\hline 
			$\bar{e}'_R\chi_L^M  H_3^-$ & $  i\frac{g}{2}Y_{H_3^-}^{ML}= i\frac{g~s_M}{2M_W c_{M}}m_{\ell}^d
            \tilde{R}_\nu \left(U_{PMNS}^{M}\right)^*$ 
			 \\
			\hline
			$\bar{e}'_L\chi_L^{Mc}  H_3^-$ & $  -i\frac{g}{2}Y_{H_3^-}^{MR}= -i\frac{g~s_M}{2M_W c_M}
			\tilde{R}_\ell m_{\ell M}^dU_{PMNS}^M$ 
			 \\
			\hline
			$\bar{e}'_R\chi_L  H_{3M}^-$ & $  -i\frac{g}{2}Y_{H_{3M}^-}^{L}= -i\frac{g~s_{2M}}
			{2M_W s_2 c_M}m_{\ell}^dU_{PMNS}$ 
			\\
			\hline 
			$\bar{e}'_R\chi_L^M  H_{3M}^-$ & $  i\frac{g}{2}Y_{H_{3M}^-}^{ML}= i\frac{g~s_{2M}}
			{2M_W s_2 c_M}m_{\ell}^d\tilde{R}_\nu \left(U_{PMNS}^{M}\right)^*$ 
			 \\
			\hline
			$\bar{e}'_L\chi_L^{Mc}  H_{3M}^-$ & $  -i\frac{g}{2}Y_{H_{3M}^-}^{MR}= -i\frac{g~s_M}{2M_W s_{2M} c_{M}}
			\tilde{R}_\ell m_{\ell M}^dU_{PMNS}^M$ 
			\\
		   \hline
			$\bar{e}'_R {e_L^M}' \phi^0_s$ & $  -i\frac{g}{2}Y_{\phi^0_s}^{ML}= -ig_{\ell s}U^{\dagger}_{\ell R}U_{\ell L}^M$ 
			\\
			\hline 
			$\bar{e}'_L {e_R^M}' \phi^0_s$ & $  -i\frac{g}{2}Y_{\phi^0_s}^{MR}= -i~g_{\ell s}U^{\dagger}_{\ell L}U_{\ell R}^M$ 
			\\
			\hline 
	\end{tabular}}
	\caption{Vertexes that contribute to the $\ell\to \ell'\gamma$ decay rates in the mass eigenstate basis. }
	\label{TableVertex}
\end{table}

\begin{equation}
\label{YuV01}
(\bar{e}'_R {e_L^M}' {\tilde{H}}_i^0)~~~~~~-i\frac{g}{2}Y^{ML}_{{\tilde{H}}_i^0}=
-i\frac{g}{2M_W}\left[\frac{\alpha_i}{s_2}m_\ell^d\tilde{R}_\ell
+\frac{\alpha_i^M}{s_{2M}}\tilde{R}_\ell m_{\ell M}^d\right],
\end{equation}
\begin{equation}
(\bar{e}'_L {e_R^M}' {\tilde{H}}_i^0)~~~~~~
-i\frac{g}{2}Y^{MR}_{{\tilde{H}}_i^0}=-i\frac{g}{2M_W}\left[\frac{\alpha_i}{s_2}m_\ell^d\tilde{R}_\ell
+\frac{\alpha_i^M}{s_{2M}}\tilde{R}_\ell m_{\ell M}^d\right],
\end{equation}

\begin{equation}
(\bar{e}'_R {e_L^M}' {H}_3^0)~~~~~~-i\frac{g}{2}Y^{ML}_{{H}_3^0}=-i\frac{g}{2M_W}\left[\frac{s_M}{c_M}m_\ell^d\tilde{R}_\ell
+\frac{s_M}{c_M}\tilde{R}_\ell m_{\ell M}^d\right],
\end{equation}
\begin{equation}
(\bar{e}'_L {e_R^M}' {H}_3^0)~~~~~~-i\frac{g}{2}Y^{MR}_{{H}_3^0}=-i\frac{g}{2M_W}\left[-\frac{s_M}{c_M}m_\ell^d\tilde{R}_\ell
-\frac{s_M}{c_M}\tilde{R}_\ell m_{\ell M}^d\right],
\end{equation}

\begin{equation}
(\bar{e}'_R {e_L^M}' {H}_{3M}^0)~~~~~~-i\frac{g}{2}Y^{ML}_{{H}_{3M}^0}=-i\frac{g}{2M_W}\left[-\frac{s_{2M}}{s_2}m_\ell^d\tilde{R}_\ell
-\frac{s_2}{s_{2M}}\tilde{R}_\ell m_{\ell M}^d\right],
\end{equation}
\begin{equation}
\label{YuV06}
(\bar{e}'_L {e_R^M}' {H}_{3M}^0)~~~~~~-i\frac{g}{2}Y^{MR}_{{H}_{3M}^0}=-i\frac{g}{2M_W}\left[\frac{s_{2M}}{s_2}m_\ell^d\tilde{R}_\ell
+\frac{s_2}{s_{2M}}\tilde{R}_\ell m_{\ell M}^d\right].
\end{equation}
Here the notations $U_{PMNS}=U_\ell^\dagger U_\nu$, which is the famous neutrino mixing PMNS matrix, $U_{PMNS}^M=U_\ell^{M\dagger} U_\nu^M$ and $\tilde{R}_{\ell (\nu)}=U_{\ell}^\dagger R_{\ell (\nu)}U_{\ell}^M$ have been used. For simplicity, we have also neglected the complex phases in $U_\ell$ and $U_\ell^M$. \\

\mathversion{bold}
\section{Phenomenology of $\mu\to e +\gamma$ decay} 
\subsection{Form-factors and $\mu\to e +\gamma$ branching ratio}
\mathversion{normal}
Before introducing the $\mu\to e +\gamma$ branching ratio, the loop
integral factors must be calculated. In this research, we take into account 
the effective charged lepton flavor changing operators arising at one-loop level
with the participation of physical Higgs scalars (which include the single and neutral charges, heavy and light ones) and $W$ gauges boson. The final result could be written as

\begin{equation}
\label{effectiveLag}
\mathcal{L}_{eff}=-4\frac{eG_F}{\sqrt{2}}m_\mu \left( A_R\bar{e}\sigma_{\mu\nu}P_R\mu
+A_L\bar{e}\sigma_{\mu\nu}P_L\mu\right) F^{\mu\nu}+H.c.
\end{equation}
Here $A_{L,R}$ are the form factors:
\begin{eqnarray}
\nonumber
A_R=&&-\sum_{{H^Q},k}\frac{M_W^2}{64\pi^2M_H^2}\left[\left(Y_H^L\right)_{\mu k}\left(Y_H^L\right)_{e k}^*G_H^Q(\lambda_k)
+\frac{m_k}{m_\mu}\left(Y_H^R\right)_{\mu k}\left(Y_H^L\right)_{e k}^*\times R_H^Q(\lambda_k)\right]\\
&&+\frac{1}{32\pi^2}\sum_{k}\left[\left(U_{W_\mu}^L\right)_{\mu k}\left(U_{W_\mu}^L\right)_{e k}^*
G_{\gamma}(\lambda_k)-\left(U_{W_\mu}^R\right)_{\mu k}\left(U_{W_\mu}^L\right)_{e k}^*
\frac{m_k}{m_\mu}R_{\gamma}(\lambda_k)\right],\label{FunAR}
\end{eqnarray}
\begin{eqnarray}
\nonumber
A_L=&&-\sum_{{H^Q},k}\frac{M_W^2}{64\pi^2M_H^2}\left[\left(Y_H^R\right)_{\mu k}\left(Y_H^R\right)_{e k}^*G_H^Q(\lambda_k)
+\frac{m_k}{m_\mu}\left(Y_H^L\right)_{\mu k}\left(Y_H^R\right)_{e k}^*R_H^Q(\lambda_k)\right]\\
&&+\frac{1}{32\pi^2}\sum_{k}\left[\left(U_{W_\mu}^R\right)_{\mu k}\left(U_{W_\mu}^R\right)_{e k}^*
G_{\gamma}(\lambda_k)-\left(U_{W_\mu}^L\right)_{\mu k}\left(U_{W_\mu}^R\right)_{e k}^*
\frac{m_k}{m_\mu}R_{\gamma}(\lambda_k)\right],\label{FunAL}
\end{eqnarray} 
where $H^Q=\phi_S^0, \tilde{H}^{0}_{i}$ $(i=1,2,3)$, $H^{0}_{3}, H^{0}_{3M}$, $H^{+}_{3}, H^{+}_{3M}$, and
$m_k$ are the masses of associated fermions that along with either $H^Q$ or $W_\mu$ to form loops. The functions 
$G_{H}^Q(x)$, $R_{H}^Q(x)$, $G_{\gamma}(x)$, and $R_{\gamma}(x)$ appearing in Eqs. (\ref{FunAR}) and (\ref{FunAL}) 
are follows
\begin{eqnarray}
G_H^Q(x)&=&-\frac{(3Q-1)x^2+5x-3Q+2}{12(x-1)^3}+\frac{1}{2}\frac{x(Qx-Q+1)}{2(x-1)^4}\log(x),\\
R_H^Q(x) &=& \frac{(2Q-1)x^2-4(Q-1)x+2Q-3}{2(x-1)^3}-\frac{Qx-(Q-1)}{(x-1)^3}\log(x),~~~~~~\\
G_\gamma(x)&=&\frac{20x^2-7x+2}{4(x-1)^3}-\frac{3}{2}\frac{x^3}{(x-1)^4}\log(x),
\\
R_\gamma(x)&=&-\frac{x^2+x-8}{2(x-1)^2}+\frac{3x(x-2)}{(x-1)^3}\log(x),
\end{eqnarray}
where we have defined $\lambda_k=m_k^2/M_{{W_\mu}({H^Q})}^2$.\\

\noindent Note that the monotonic functions $G_\gamma(x)$ and $R_\gamma(x)$, which are defined for
$x$ variable varying in the interval $[0,+\infty)$, have been introduced in some researches so far, for instance \cite{Alonso:2012ji,Dinh:2012bp}. At some specific points, such as $x=0,~1$ and $x$ tends to infinity, $G_\gamma(x)$ obtains the limited values as $-1/2$, $-3/8$ and $0$, respectively. Similarly, we also have $R_\gamma(x\rightarrow 0)=4$, $R_\gamma(x\rightarrow 1)=3/2$ and $R_\gamma(x\rightarrow\infty)=-1/2$. Compare with the form-factors
for Higgs scalar one-loop have been used in some previous publications \cite{Akeroyd:2009nu, Dinh:2012bp}, $G_{H}^Q(x)$  has better expression {\footnote{$G_{H}^Q(x)$ introduced the current research is valid for any $x$ in the interval $[0,+\infty)$, while the previous calculations are 
applied only for infinitesimal $\lambda_k$. When $x$ tends to zero, functions $G_{H}^Q(x)$ tends to $1/6-Q/4$, that is consistent to the results obtained in \cite{Akeroyd:2009nu, Dinh:2012bp}.}}, and $R_{H}^Q(x)$, as far as my knowledge, has not been given so far.\\

\noindent The branching ratio of $\mu\to e+\gamma$ decay is easily obtained as   
\begin{equation}
\label{mutoegammarate}
{\rm Br}{(\mu\to e+\gamma)}=384\pi^2(4\pi\alpha_{em})\left(|A_R|^2+|A_L|^2\right),
\end{equation}
where $\alpha_{em}=1/137$ is the fine-structure constant. 
\mathversion{bold}
\subsection{Numerical analysis of the $\mu\to e +\gamma$ decay}
\mathversion{normal}
We perform in this section the numerical analysis of $\mu\to e +\gamma$ branching ratio
using the current experimental data and expected sensitivity of the future experiments. Apparently, taking into account at the same time all the possible contributions to the
process would not a good strategy to give detail understanding the role of each kind of the diagrams. Thus, we separately consider the contributions of the one-loops diagrams with virtual $W$ gauge boson, neutral and singly charged Higgs scalars  to the ratio. For simplicity in further numerical discussion in the later part, we suppose that three heavy neutrinos  possess equal masses denoted as $m_\chi^M$. Similarly, three mirror charged lepton masses are $m_\ell^M$.\\ 

\noindent The LFV vertexes involving neutrinos and W boson taken upto $R_{\nu(\ell)}$ first order are the three firsts listed in the table \ref{TableVertex}. In fact, the contribution of light neutrinos to the $\mu\to e +\gamma$ decay are extremely small, which could be easily seen from eq. (\ref{FunAR}). For light neutrino masses are at sub-eV order or less, and $M_W=80$ GeV, $\lambda_k=m_k^2/M_W^2\approx 0$, that leads to (see eq. (\ref{FunAR}))
\begin{equation}
\sum_{k=1}^3\left(U_{W_\mu}^L\right)_{\mu k}\left(U_{W_\mu}^L\right)_{e k}^*
G_{\gamma}(\lambda_k)\approx \sum_{k=1}^3\left(U_{W_\mu}^L\right)_{\mu k}\left(U_{W_\mu}^L\right)_{e k}^*
G_{\gamma}(0)=\left(U_{W_\mu}^{L\dagger}U_{W_\mu}^{L}\right)_{e\mu}G_{\gamma}(0)\approx 0,
\end{equation}
due to the unitarity of PMNS matrix. Moreover, contribution from the interference term 
in $A_R$ is strongly suppressed by the factor $m_k/m_\mu\sim 10^{-1}~{\rm eV}/10^2 ~{\rm MeV}\sim 10^{-9}$.\\

\begin{figure}[t]
\begin{center}
\includegraphics[width=12cm,height=8cm]{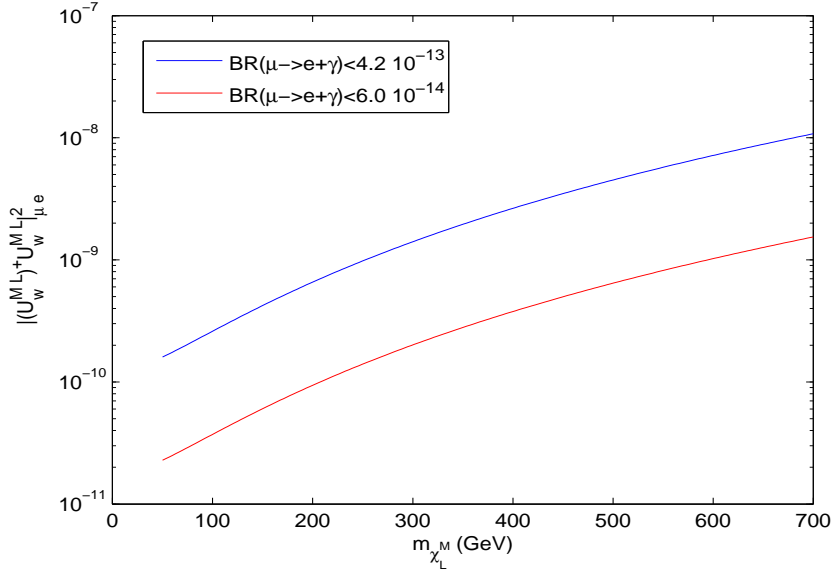}
\caption{Constraint on $\left(U_{W}^{ML\dagger}U_{W}^{ML}\right)_{e\mu}$
 by $\mu\to e\gamma$ decay from current (blue line) and future expected (red line) sensitivities as function of heavy neutrino mass $m_\chi^M$.}
\label{GaugePlot}
\end{center}
\end{figure}

\noindent Constraint on the interaction strength between heavy neutrinos and W gauge boson by the lepton flavour violation decay $\mu\to e\gamma$ is given in figure \ref{GaugePlot}. The blue and red lines correspond to the constraints obtained using whether the current upper bound or the designed sensitivity of the future experiment, respectively. The results, apparently, are less meaningful comparing with the limits, which could be directly derived from what we have already known about light neutrino masses. One have
%
\begin{eqnarray}
{\tilde{m}}_\nu=\frac{(m_\nu^D)^2}{M_R}\sim 10^{-10} {\rm~GeV}\Rightarrow 
R_\nu=\frac{m_\nu^D}{M_R}\sim 10^{-5}\sqrt{\frac{1{\rm GeV}}{M_R}}, 
\end{eqnarray}
thus $|R_\nu|^2\sim 10^{-12}$ for $M_R\sim 100 {\rm ~GeV}$ that is at least 7 (6) orders smaller than the constraints obtained from figure \ref{GaugePlot}:
$\left|U_{W}^{ML\dagger}U_{W}^{ML}\right|_{e\mu}\sim |R_\nu|^2< 10^{-5}~(3\times 10^{-6})$ corresponding to the current (near future) experimental sensitivities.\\
\begin{figure}[t]
\begin{center}
\includegraphics[width=12cm,height=8cm]{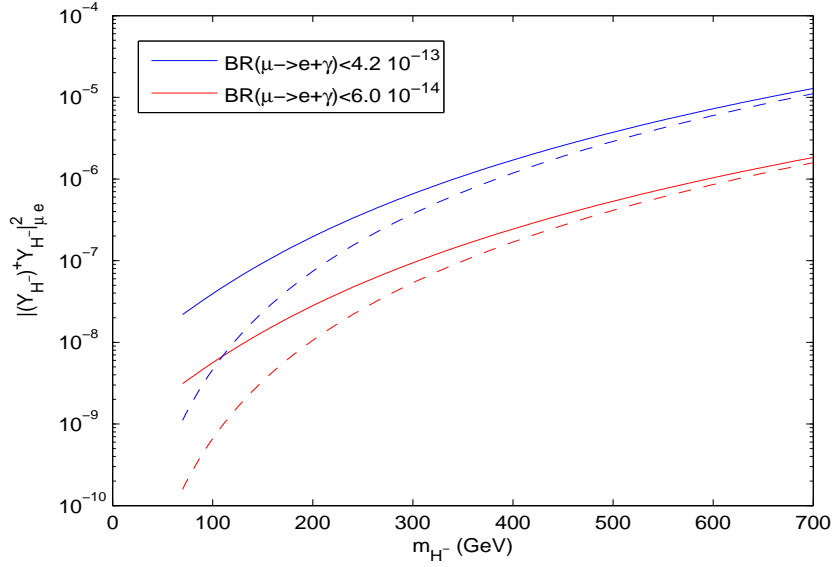}
\caption{Constraint on Yukawa couplings as function of physical singly charged Higgs scalar mass by $\mu\to e\gamma$ decay if only left or right
sector is taken into account, for new neutrino masses are $150$ GeV (solid lines) and $10^{-9}$ GeV (dash lines).}
\label{SingleH1Plot}
\end{center}
\end{figure}
%

\noindent Unlike the earlier considered case, contributions of diagrams with participation of virtual Higgs scalars (for both neutral and singly charged ones) contain fully two terms given in (\ref{FunAR}), (\ref{FunAL}). The interference term is no longer suppressed but dominated over the first term due to the large masses of accompanying particles with the physical scalar, which are heavy neutrinos or new charged leptons depending on kind of charge carried by the scalar.
For heavy neutrino mass $m_{\chi^M_L}$ and new charged lepton mass $m_\ell^M$ about hundreds GeV, the ratio $m_k/m_\mu\sim 100 {\rm ~GeV}/100 {\rm~MeV}\sim 10^{3}$, that is also the dominated factor of the second term in comparison with the first term.\\

\noindent We show in figure \ref{SingleH1Plot} constraint on the Yukawa coupling only if the first term in (\ref{FunAR}), (\ref{FunAL}) are taken into account for singly charged scalar cases. The stringency obtained in this case on the magnitude of the couplings is almost the same as the previous consideration of W gauge boson and heavy neutrino couplings, and thus does not provide any new meaning. Our study shows that similar results are obtained for the case of neutral Higgs scalars.\\
%
\begin{figure}
\begin{center}
\begin{tabular}{cc}
\includegraphics[width=7.5cm,height=6.5cm]{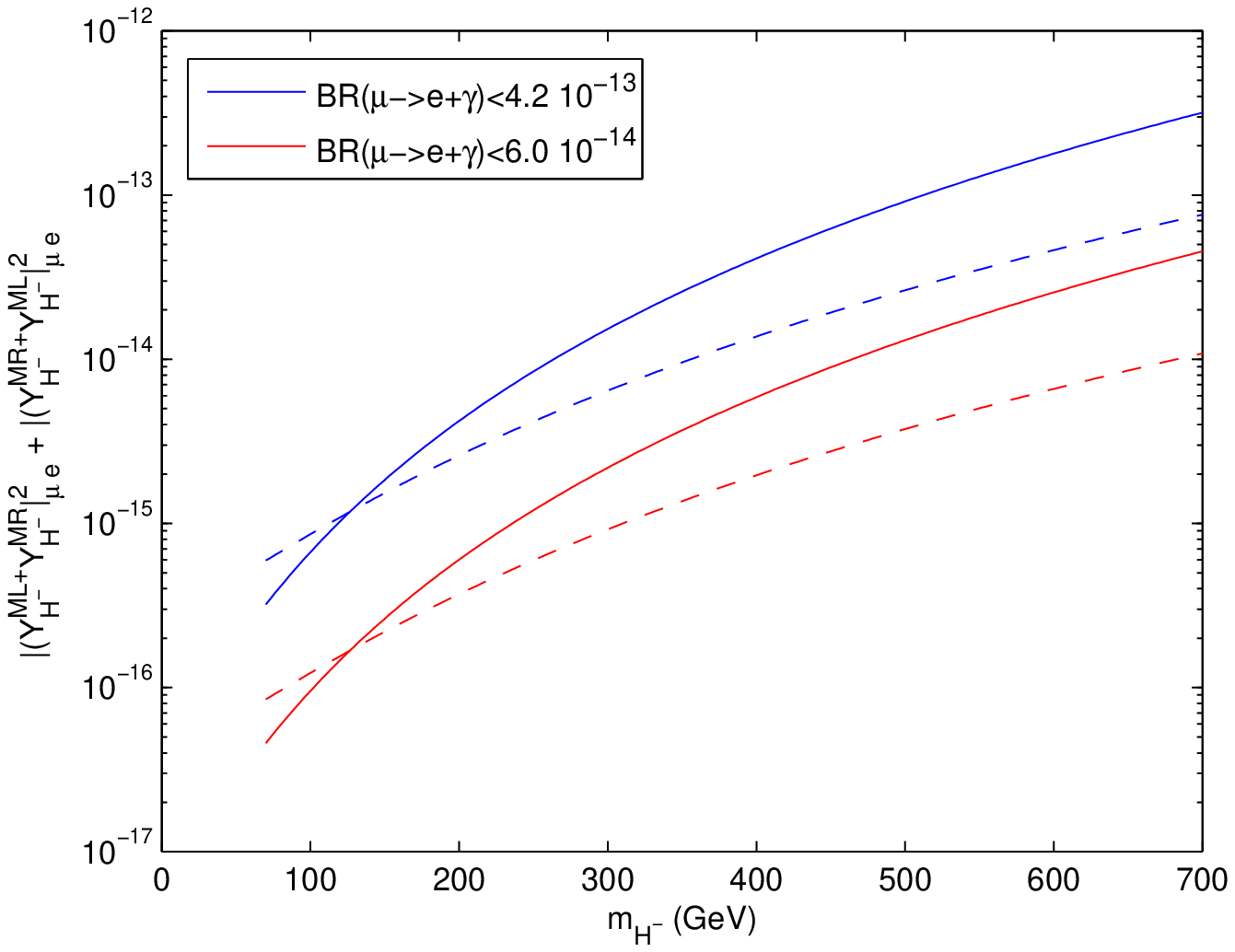} &
\includegraphics[width=7.5cm,height=6.5cm]{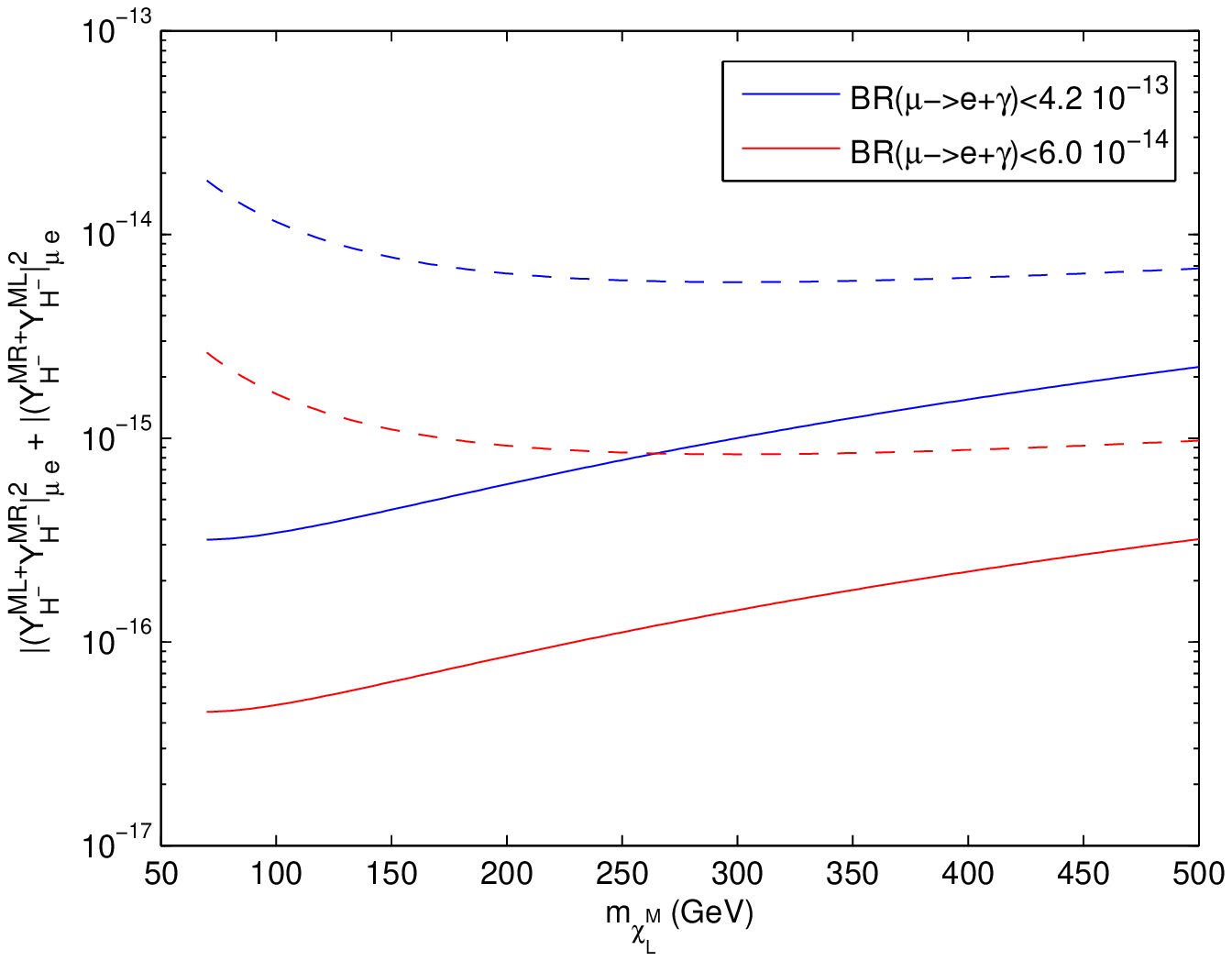}\\
\end{tabular}
\caption{Upper bounds on Yukawa couplings by $\mu\to e\gamma$ decay as functions of: i, Singly charged Higgs mass (left-panel) for $m_{\chi^M_L}=80~(200)~{\rm GeV}$, solid (dash) lines; ii, Heavy neutrino masses (right-panel) for $m_{H^-}=70~(300)~{\rm GeV}$, solid (dash) lines.}
\label{Contri_H1YrYL}
\end{center}
\end{figure}

\noindent Contributions of the first terms in the expressions of the form-factors $A_L$ and $A_R$ are less important, that is because they are strongly suppressed by the second terms as explained in a previous paragraph. The figure \ref{Contri_H1YrYL} describes the upper constraints on the relevant Yukawa couplings as functions of Higgs scalar mass (left-panel) and heavy neutrino mass (right-panel), which are varied from about hundred GeV to several hundreds GeV. The constraints are about six orders more stringent than the values of those obtained from previous figure. That, one again, proves the dominated contribution of the interference term and is also consistent with the illustration given somewhere above. The left plot shows lines, which have shapes of monotonically increasing functions, therefore the most stringent constraints on $|(Y_{H^-}^{ML})^\dagger Y_{H^-}^{MR}|^2_{\mu e}+|(Y_{H^-}^{MR})^\dagger Y_{H^-}^{ML}|^2_{\mu e}$ are at the initial points of the lines with $m_{H^-}\sim 100~{\rm GeV}$. The results read:
\begin{eqnarray}
|(Y_{H^-}^{ML})^\dagger Y_{H^-}^{MR}|^2_{\mu e}+|(Y_{H^-}^{MR})^\dagger Y_{H^-}^{ML}|^2_{\mu e}\lesssim 3.0\times 10^{-16}(6.0\times 10^{-16})~{\rm for}~m_{\chi^M_L}=80~(200)~{\rm GeV},
\end{eqnarray}
using the present experimental upper bound; and 
\begin{eqnarray}
|(Y_{H^-}^{ML})^\dagger Y_{H^-}^{MR}|^2_{\mu e}+|(Y_{H^-}^{MR})^\dagger Y_{H^-}^{ML}|^2_{\mu e}\lesssim 4.0\times 10^{-17}(8.0\times 10^{-17})~{\rm for}~m_{\chi^M_L}=80~(200)~{\rm GeV},
\end{eqnarray}
using the expected upper bound in the future.\\

\noindent Note that figure \ref{Contri_H1YrYL} is common for both $H_3^{-}$ and $H_{3M}^{-}$, however the specific forms of Yukawa couplings involving them (see table \ref{TableVertex}) will decide how sensitive they are with the $\mu\to e\gamma$ decay experiments. 
The involved couplings depend on lager number of new parameters, where most of them are unknown, thus it is too difficult if not want to say impossible to make a detail analysis. In this research, we try to roughly make estimation on the sensitivities of 
the $\mu\to e\gamma$ decay with the present and future experiments, using the known data and supposing that the model is functioning at the electroweak scale.\\

\noindent Let us make a numerical estimation. As have been explained in the earlier part that $R_\nu\sim 10^{-5}\sqrt{\frac{1{\rm GeV}}{M_R}}\sim 10^{-6}$. In the same way, we also have $R_\ell\sim 10^{-5}\sqrt{\frac{1{\rm GeV~M_R}}{M_{\ell M}^2}}\sim 10^{-6}$. It is reasonable to estimate $\tilde{R}_{\ell (\nu)}=U_{\ell}^\dagger R_{\ell (\nu)}U_{\ell}^M\sim 10^{-6}$, at the same order as $R_\ell$ and $R_\nu$, since basis transformation matrices $U_{\ell}$ and $U_{\ell}^M$ are normalized. Furthermore, we take heavy neutrino and mirror charged lepton masses about $100$ GeV. The Yukawa coupling depended factor of the branching ratio in eq. (\ref{mutoegammarate}) reads:
\begin{eqnarray}
\left|Y_{H^-_{3M}}^{L\dagger}Y_{H^-_{3M}}^{R}\right|^2_{21}&+&\left|Y_{H^-_{3M}}^{R\dagger}Y_{H^-_{3M}}^{L}\right|^2_{21}=\left(\frac{s_M}{s_2c_M^2}\right)^2
\frac{\left|m_\ell^d\tilde{R}_\nu m_{\ell M}^d\tilde{R}_\ell\right|^2_{21} + 
\left|\tilde{R}_\ell m_{\ell M}^d\tilde{R}_\nu m_{\ell}^d\right|^2_{21}}{M_W^4}~\\
\label{limitcalH1}
&&\sim 10^{-24}\left(\frac{s_M}{s_2c_M^2}\right)^2\times \left(\frac{3m_\mu m_\ell^M}{M_W^2}\right)^2\sim
2.2\times 10^{-29}\left(\frac{s_M}{s_2c_M^2}\right)^2,
\end{eqnarray}
here we have ignored the second term (proportional to $m_e$), which is strongly suppressed by the first term (proportional to $m_\mu$). This result means that 
$\mu\to e\gamma$ branching ratio might be within the sensitive limit of future experiment 
only if $s_2$ and $c_M$ are both equal or smaller than $0.01$. For $H^-_{3}$ case, the factor $\left(\frac{s_M}{s_2c_M^2}\right)^2$ in eq. (\ref{limitcalH1}) is replaced by $\left(\frac{s_M^2}{c_M^2}\right)^2$, thus the quantity $\left|Y_{H^-_{3}}^{L\dagger}Y_{H^-_{3}}^{R}\right|^2_{21}$ would reach order of $10^{-17}$, if $c_M\lesssim 0.001$.\\
\begin{figure}
\begin{center}
\begin{tabular}{cc}
\includegraphics[width=7.5cm,height=6.5cm]{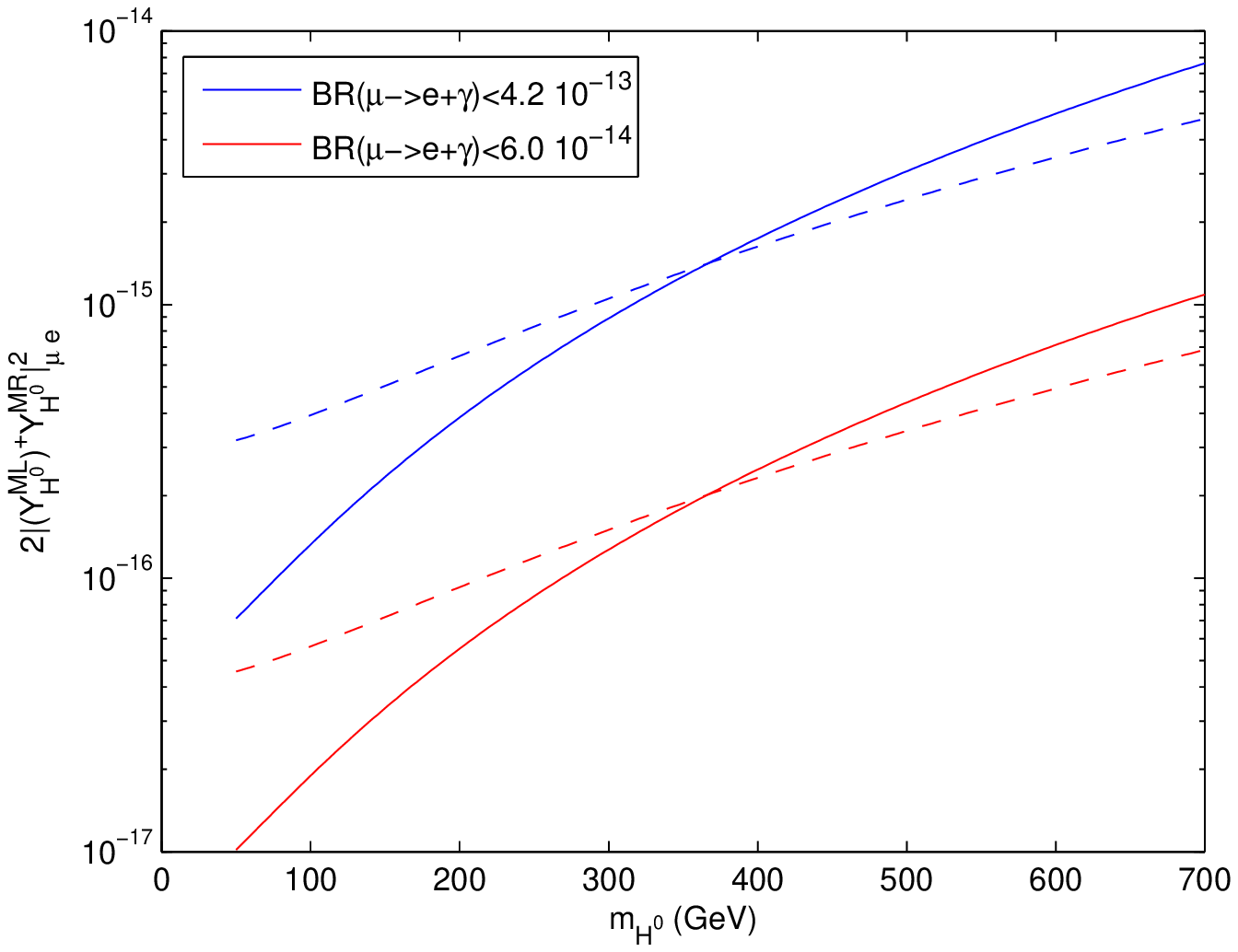} &
\includegraphics[width=7.5cm,height=6.5cm]{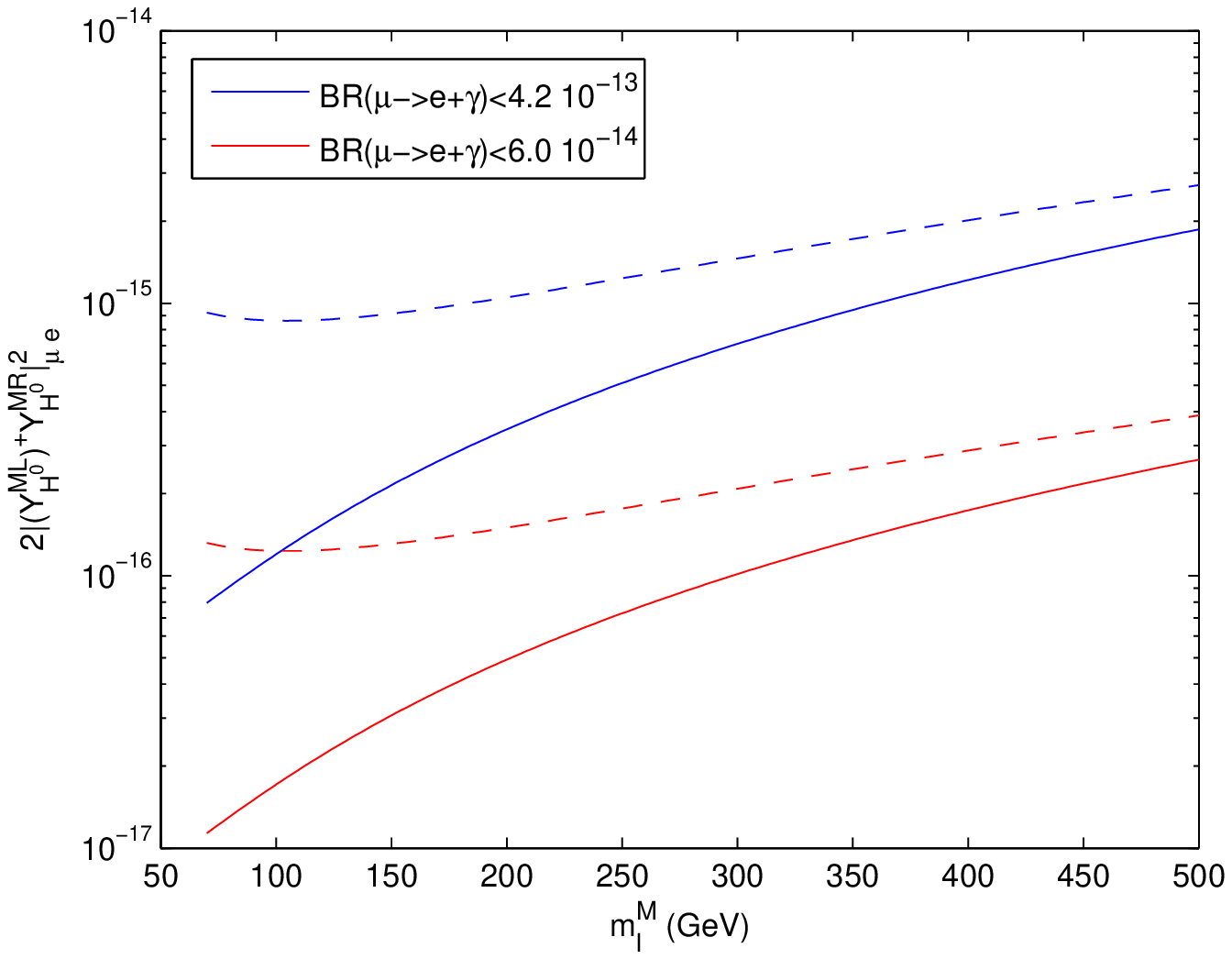}\\
\end{tabular}
\caption{Upper bounds on Yukawa couplings involving $H^0$ ($H^0_{3},~H^0_{3M},~\tilde{H}^0_i$, $(i=1,2,3)$), by $\mu\to e\gamma$ decay as functions of: i, Neutral Higgs scalar mass (left-panel) for $m_{\ell}^M=80~(200)~{\rm GeV}$, solid (dash) lines; ii, Mirror charged lepton masses (right-panel) for $m_{H^0}=70~(300)~{\rm GeV}$, solid (dash) lines.}
\label{Contri_H0YrYL}
\end{center}
\end{figure}

\noindent Due to the enormous disparity between their masses, we anlysize physical phenomenology for the light and heavy neutral Higgs scalars separately. Note that the first terms of Yukawa couplings, listed from eq. (\ref{YuV01}) to eq. (\ref{YuV06}), are proportional to 
normal charged lepton masses, thus are ignorable in comparison with the others, which are
proportional to heavier masses of the mirror leptons. We show in figure \ref{Contri_H0YrYL} the constraints on relevant Yukawa couplings as functions of Higgs scalar mass (left-panel) and heavy charged lepton mass (right-panel), respectively. One easily realize that the heavier the Higgs boson and mirror charged lepton masses are the less stringent constraints are given by $\mu\to e\gamma$ decay. Moreover, compare with the singly charged Higgs scalar cases, constraints obtained in this case are little more stringent.
One easily obtains from the left panel of figure \ref{Contri_H0YrYL}, that:
\begin{eqnarray}
2|(Y_{H^0}^{ML})^\dagger Y_{H^0}^{MR}|^2_{\mu e}\lesssim 7.0\times 10^{-17}(3.0\times 10^{-16})~{\rm for}~m_{\ell}^M=80~(200)~{\rm GeV},
\end{eqnarray}
with the present experimental upper bound; and 
\begin{eqnarray}
2|(Y_{H^0}^{ML})^\dagger Y_{H^0}^{MR}|^2_{\mu e}\lesssim 1.0\times 10^{-17}(4.5\times 10^{-17})~{\rm for}~m_{\ell}^M=80~(200)~{\rm GeV},
\end{eqnarray}
with the future expected upper bound.\\

\noindent Carry on similar estimation as the previous part, the results are:
\begin{eqnarray}
\label{limitcalH0}
2\left|Y_{H^0}^{L\dagger}Y_{H^0}^{R}\right|^2_{21}=2\alpha^4\times
\frac{\left|\tilde{R}_\ell (m_{\ell M}^d)^2\tilde{R}_\ell\right|^2_{21}}{M_W^4}\sim 4.4\times 10^{-23}\alpha^4,
\end{eqnarray}
where $\alpha$ stands for $\frac{\alpha_i}{s_2}$, $\frac{s_M}{c_M}$ or $\frac{s_2}{s_{2M}}$, corresponding to $\tilde{H}^0_i,~(i=1,2,3)$, $H^0_3$ or $H^0_{3M}$, respectively. Compare with eq. (\ref{limitcalH1}), this result is about six orders higher  (more sensitive) because light charged lepton diagonal matrix $m_{\ell}^d$ in earlier formula has been replaced by the heavy one $m_{\ell M}^d$. In the $H_3^0$ case, for $c_M=0.01$, the factor $2\left|Y_{H^0}^{L\dagger}Y_{H^0}^{R}\right|^2_{21}\sim 4.4\times 10^{-15}$, which is about two order higher than the upper constraint of that if the decay signal would not be probed by the future experiment . In fact, our calculation show that the factor might be lager than $10^{-17}$ with $c_M\lesssim 0.03$. We have the same conclusions if $s_2\sim 0.01$, $\alpha_i$ is not small (for $\tilde{H}_i^0$) and $s_{2M}\sim 0.01$, $s_2$ is not small (for $H_{3M}^0$). Note that contributions to the decay rate by $\tilde{H}_i^0$ and $H_{3M}^0$ channels are not likely to be sensitive at the same time, due to their 
dependents on $s_2$ are in opposite way. \\
%
\begin{figure}
\begin{center}
\begin{tabular}{cc}
\includegraphics[width=7.5cm,height=6.5cm]{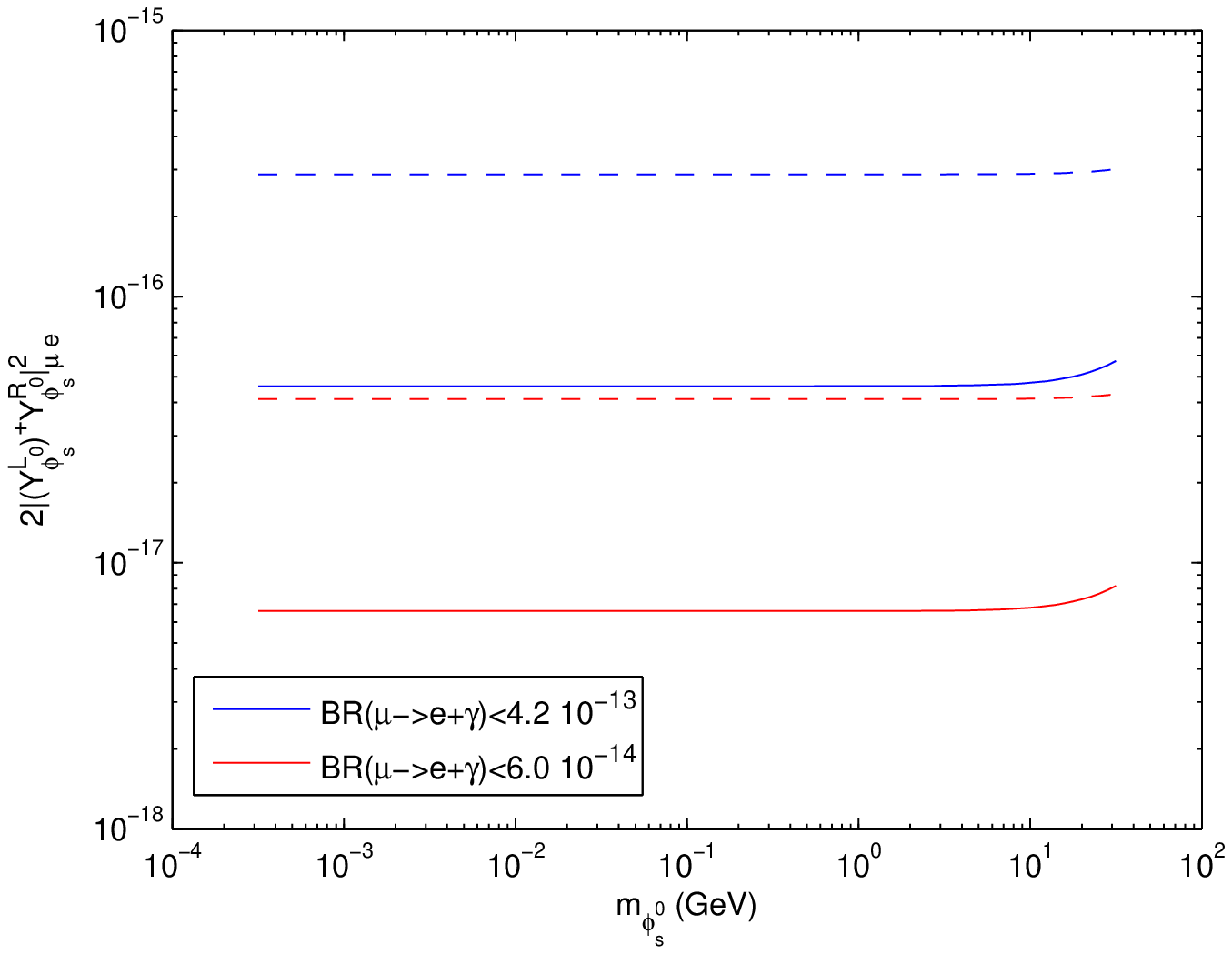} &
\includegraphics[width=7.5cm,height=6.5cm]{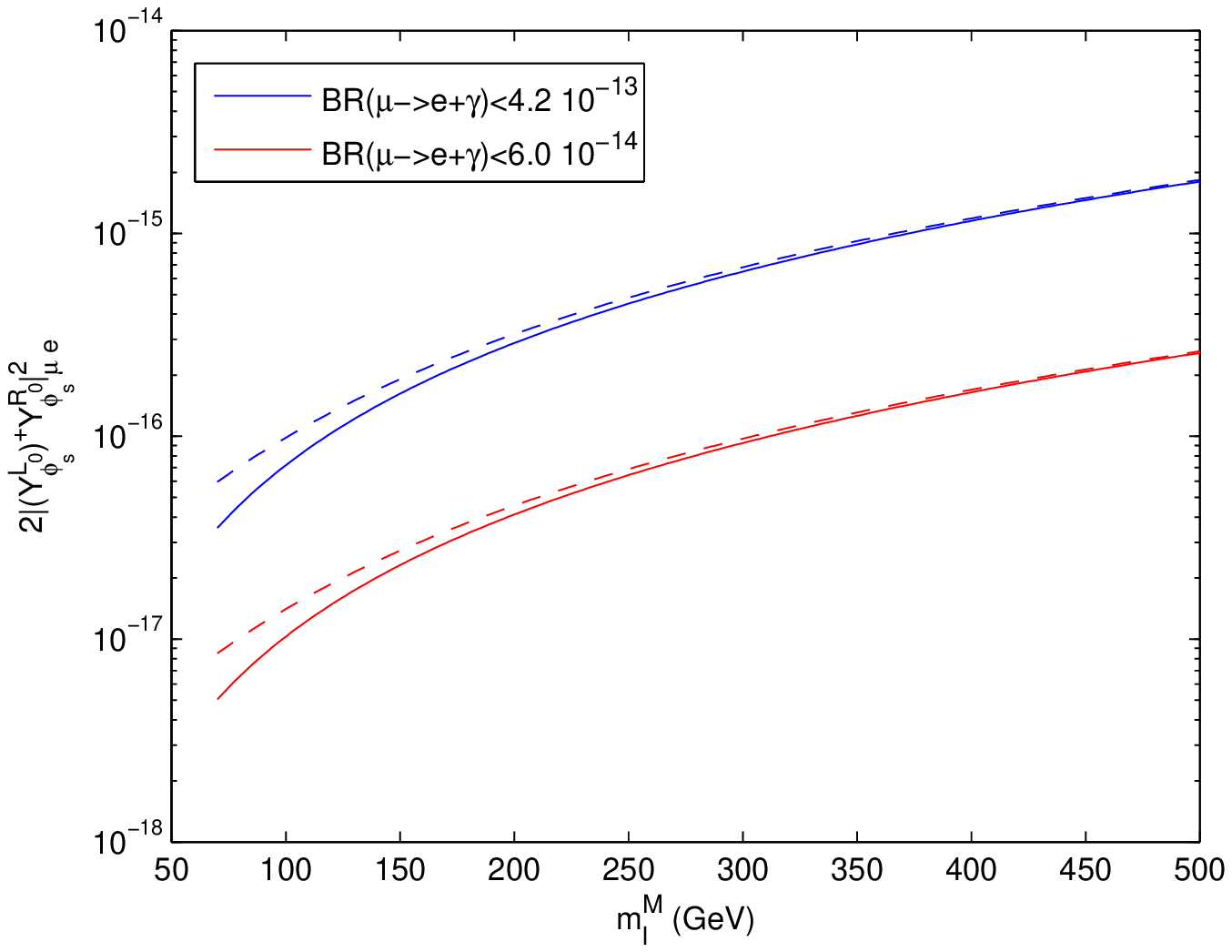}\\
\end{tabular}
\caption{The dependent of $2|(Y_{\phi_s^0}^{L})^\dagger Y_{\phi_s^0}^{R}|^2_{\mu e}$ upper limits by current (future expected) sensitivities, corresponding to blue (red) lines, on: Light neutral Higgs scalar mass (left-panel) for $m_{\ell}^M=80~(200)~{\rm GeV}$, solid (dash) lines; ii, Mirror charged lepton masses (right-panel) for $m_{\phi^0_s}=10^{-3}~(50)~{\rm GeV}$, solid (dash) lines. }
\label{Contri_Phi0YrYL}
\end{center}
\end{figure}

\noindent Constraints on the Yukawa couplings involving light Higgs scalar are shown in figure \ref{Contri_Phi0YrYL}, in which the scalar mass is varied in a large range from $\rm KeV$
to more than $10{\rm GeV}$ (left-panel). The figure shows that constrained stringency on the Yukawa couplings does not change as increasing of $m_{\phi^0_s}$ until about $10{\rm GeV}$, then slowly decreases. Precise values of upper limits obtained from red and blue lines, respectively, are:
\begin{equation}
\label{litmit01}
2|(Y_{\phi_s^0}^{L})^\dagger Y_{\phi_s^0}^{R}|^2_{\mu e}\lesssim 4.7\times 10^{-17}(3.0\times 10^{-16})~{\rm for}~m_{\ell}^M=80~(200)~{\rm GeV},
\end{equation}
\begin{equation}
\label{litmit02}
2|(Y_{\phi_s^0}^{L})^\dagger Y_{\phi_s^0}^{R}|^2_{\mu e}\lesssim 6.5\times 10^{-18}(4.2\times 10^{-17})~{\rm for}~m_{\ell}^M=80~(200)~{\rm GeV}.
\end{equation}
It can be recasted from eqs. (\ref{litmit01}) and (\ref{litmit02}) respectively into
upper bound on $|g_{\ell s}|$ as:
\begin{equation}
\label{litmitg01}
|g_{\ell s}|\lesssim 2.3\times 10^{-5}(3.6\times 10^{-5})~{\rm for}~m_{\ell}^M=80~(200)~{\rm GeV},
\end{equation}
\begin{equation}
\label{litmitg02}
|g_{\ell s}|\lesssim 1.4\times 10^{-5}(2.3\times 10^{-5})~{\rm for}~m_{\ell}^M=80~(200)~{\rm GeV}.
\end{equation}
Thus $|g_{\ell s}|$ upper limit is estimated with current experimental data to be at order $10^{-5}$, it might be improved little more if next generation of $\mu\to e\gamma$
experiment would not probe any signal. Moretheless the constraint is still in the same scale.\\

\noindent As one can see in the right-panel of figure \ref{Contri_Phi0YrYL}, constraints on the Yukawa couplings become less and less stringent as mirror charged lepton mass goes up. However, the shape does not effect strongly on $|g_{\ell s}|$ upper bound. Let us make a simple evaluation. For $m_\ell^{M}=500$ GeV, which corresponds to the highest or the end point of each line, one has $2|(Y_{\phi_s^0}^{L})^\dagger Y_{\phi_s^0}^{R}|^2_{\mu e}\lesssim 1.8\times 10^{-15}(2.6\times 10^{-16})$, with ignorable dependence on the light Higgs scalar mass. The result leads to $|g_{\ell s}|\lesssim 5.7\times 10^{-5}(3.5\times 10^{-5})$, which are rather close to what have been obtained in (\ref{litmitg01}) and (\ref{litmitg02}). This fact means energy scale of $v_S$ (thus, also the light Higgs mass $m_{\phi^0_s}$) is order GeV or more, if one wants to keep neutrino mass generation scale at EW and ensure the light active neutrino mass ($\tilde{m}_\nu\approx (m_\nu^D)^2/M_R=(g_{\ell s} v_S)^2/M_R$) about su-eV as results have been experimentally observed.
\section{\label{conl} Conclusion}
In this research, we have performed a numerical analysis for the $\mu\to e\gamma$ 
decay in a scenario of the EW-scale non-sterile right-handed neutrino model, which is accommodated with the $125$ GeV SM-like scalar discovery. The decay is suggested to 
occur at one-loop diagrams, formed by neutrinos (heavy and light) accompanying with 
W-boson or singly charged scalars, and light or heavy neutral scalars with charged leptons. It has been shown that the contribution provided by neutrino and W-boson loop 
channels give trivial upper constraint, which is about six orders less stringent than the limit of that directly derived from neutrino mass currently known data. For the case  particles running inside the loops are light scalar and mirror charged leptons, upper bound for the Yukawa couplings $|g_{\ell s}|$ roughly obtained after comparing theoretical prediction and experimental results is some number of order $10^{-5}$. Consequently 
that brings the singlet vacuum expectation value upto magnitude of few GeVs or
few ten GeVs, if the RH neutrino mass is managed within the electroweak scale. The 
research has also demonstrated that the branching ratio of $\mu\to e\gamma$ decay might be large enough to reach expected sensitivity, ${\rm Br}(\mu\to e\gamma)\le 6.0\times 10^{-14}$, of the upgraded MEG experiment, if one of the two particles running in the loop is the heavy scalar with neutral or single charge. For instance, as one of the most promising possibilities, magnitude of the branching ratios might be within the detectable range with $c_M\le 0.03$ for the case neutral scalar $H_3^0$ participating in the process. 
 
\section*{Acknowledgments}
	
This research is funded by Vietnam National Foundation for Science and Technology Development (NAFOSTED) under grant number 103.01-2019.307.
	
\bibliography{combine}

\begin{thebibliography}{10}

\bibitem{Hung:2006ap}
P.~Q. Hung, ``{A Model of electroweak-scale right-handed neutrino mass},'' {\em
  Phys. Lett. B}, vol.~649, pp.~275--279, 2007.

\bibitem{Minkowski:1977sc}
P.~Minkowski, ``{$\mu \to e\gamma$ at a Rate of One Out of $10^{9}$ Muon
  Decays?},'' {\em Phys. Lett. B}, vol.~67, pp.~421--428, 1977.

\bibitem{Gell-Mann:1979vob}
M.~Gell-Mann, P.~Ramond, and R.~Slansky, ``{Complex Spinors and Unified
  Theories},'' {\em Conf. Proc. C}, vol.~790927, pp.~315--321, 1979.

\bibitem{Yanagida:1979as}
T.~Yanagida, ``{Horizontal gauge symmetry and masses of neutrinos},'' {\em
  Conf. Proc. C}, vol.~7902131, pp.~95--99, 1979.

\bibitem{Mohapatra:1979ia}
R.~N. Mohapatra and G.~Senjanovic, ``{Neutrino Mass and Spontaneous Parity
  Nonconservation},'' {\em Phys. Rev. Lett.}, vol.~44, p.~912, 1980.

\bibitem{Hoang:2013jfa}
V.~Hoang, P.~Q. Hung, and A.~S. Kamat, ``{Electroweak precision constraints on
  the electroweak-scale right-handed neutrino model},'' {\em Nucl. Phys. B},
  vol.~877, pp.~190--232, 2013.

\bibitem{Aad:2012tfa}
G.~Aad {\em et~al.}, ``{Observation of a new particle in the search for the
  Standard Model Higgs boson with the ATLAS detector at the LHC},'' {\em
  Phys.Lett.}, vol.~B716, pp.~1--29, 2012.

\bibitem{Chatrchyan:2012ufa}
S.~Chatrchyan {\em et~al.}, ``{Observation of a new boson at a mass of 125 GeV
  with the CMS experiment at the LHC},'' {\em Phys.Lett.}, vol.~B716,
  pp.~30--61, 2012.

\bibitem{Hoang:2014pda}
V.~Hoang, P.~Q. Hung, and A.~S. Kamat, ``{Non-sterile electroweak-scale
  right-handed neutrinos and the dual nature of the 125-GeV scalar},'' {\em
  Nucl. Phys. B}, vol.~896, pp.~611--656, 2015.

\bibitem{Petcov:1976ff}
S.~T. Petcov, ``{The Processes $\mu \rightarrow e + \gamma, \mu \rightarrow e +
  \overline{e}, \nu' \rightarrow \nu + \gamma$ in the Weinberg-Salam Model with
  Neutrino Mixing},'' {\em Sov. J. Nucl. Phys.}, vol.~25, p.~340, 1977.
\newblock [Erratum: Sov.J.Nucl.Phys. 25, 698 (1977), Erratum: Yad.Fiz. 25, 1336
  (1977)].

\bibitem{Bilenky:1977du}
S.~M. Bilenky, S.~T. Petcov, and B.~Pontecorvo, ``{Lepton Mixing, mu
  --\ensuremath{>} e + gamma Decay and Neutrino Oscillations},'' {\em Phys.
  Lett. B}, vol.~67, p.~309, 1977.

\bibitem{Cheng:1980tp}
T.~P. Cheng and L.-F. Li, ``{$\mu \to e \gamma$ in Theories With Dirac and
  Majorana Neutrino Mass Terms},'' {\em Phys. Rev. Lett.}, vol.~45, p.~1908,
  1980.

\bibitem{TheMEG:2016wtm}
A.~M. Baldini {\em et~al.}, ``{Search for the lepton flavour violating decay
  $\mu ^+ \rightarrow \mathrm {e}^+ \gamma $ with the full dataset of the MEG
  experiment},'' {\em Eur. Phys. J.}, vol.~C76, no.~8, p.~434, 2016.

\bibitem{MEGII:2021fah}
A.~M. Baldini {\em et~al.}, ``{The Search for $\mu^+\to e^+ \gamma$ with
  10$^{-14}$ Sensitivity: the Upgrade of the MEG Experiment},'' 7 2021.

\bibitem{Renga:2014xra}
F.~Renga, ``{Latest results of MEG and status of MEG-II},'' in {\em {20th
  International Conference on Particles and Nuclei}}, 9 2014.

\bibitem{Hisano:1995cp}
J.~Hisano, T.~Moroi, K.~Tobe, and M.~Yamaguchi, ``{Lepton flavor violation via
  right-handed neutrino Yukawa couplings in supersymmetric standard model},''
  {\em Phys. Rev. D}, vol.~53, pp.~2442--2459, 1996.

\bibitem{Alonso:2012ji}
R.~Alonso, M.~Dhen, M.~B. Gavela, and T.~Hambye, ``{Muon conversion to electron
  in nuclei in type-I seesaw models},'' {\em JHEP}, vol.~01, p.~118, 2013.

\bibitem{Kakizaki:2003jk}
M.~Kakizaki, Y.~Ogura, and F.~Shima, ``{Lepton flavor violation in the triplet
  Higgs model},'' {\em Phys. Lett. B}, vol.~566, pp.~210--216, 2003.

\bibitem{Akeroyd:2009nu}
A.~G. Akeroyd, M.~Aoki, and H.~Sugiyama, ``{Lepton Flavour Violating Decays tau
  ---\ensuremath{>} anti-l ll and mu ---\ensuremath{>} e gamma in the Higgs
  Triplet Model},'' {\em Phys. Rev. D}, vol.~79, p.~113010, 2009.

\bibitem{Leontaris:1985qc}
G.~K. Leontaris, K.~Tamvakis, and J.~D. Vergados, ``{Lepton and Family Number
  Violation From Exotic Scalars},'' {\em Phys. Lett. B}, vol.~162,
  pp.~153--159, 1985.

\bibitem{Raidal:1997hq}
M.~Raidal and A.~Santamaria, ``{Muon electron conversion in nuclei versus mu
  ---\ensuremath{>} e gamma: An Effective field theory point of view},'' {\em
  Phys. Lett. B}, vol.~421, pp.~250--258, 1998.

\bibitem{Ma:2000xh}
E.~Ma, M.~Raidal, and U.~Sarkar, ``{Phenomenology of the neutrino mass giving
  Higgs triplet and the low-energy seesaw violation of lepton number},'' {\em
  Nucl. Phys. B}, vol.~615, pp.~313--330, 2001.

\bibitem{Petcov:1982en}
S.~T. Petcov, ``{Remarks on the Zee Model of Neutrino Mixing (mu
  ---\ensuremath{>} e gamma, Heavy Neutrino ---\ensuremath{>} Light Neutrino
  gamma, etc.)},'' {\em Phys. Lett. B}, vol.~115, pp.~401--406, 1982.

\bibitem{Abada:2007ux}
A.~Abada, C.~Biggio, F.~Bonnet, M.~B. Gavela, and T.~Hambye, ``{Low energy
  effects of neutrino masses},'' {\em JHEP}, vol.~12, p.~061, 2007.

\bibitem{Hung:2007ez}
P.~Q. Hung, ``{Electroweak-scale mirror fermions, mu ---\ensuremath{>} e gamma
  and tau ---\ensuremath{>} mu gamma},'' {\em Phys. Lett. B}, vol.~659,
  pp.~585--592, 2008.

\bibitem{Hung:2015hra}
P.~Q. Hung, T.~Le, V.~Q. Tran, and T.-C. Yuan, ``{Lepton Flavor Violating
  Radiative Decays in EW-Scale $\nu_R$ Model: An Update},'' {\em JHEP},
  vol.~12, p.~169, 2015.

\bibitem{Chang:2017vzi}
C.-F. Chang, P.~Q. Hung, C.~S. Nugroho, V.~Q. Tran, and T.-C. Yuan, ``{Electron
  Electric Dipole Moment in Mirror Fermion Model with Electroweak Scale
  Non-sterile Right-handed Neutrinos},'' {\em Nucl. Phys. B}, vol.~928,
  pp.~21--37, 2018.

\bibitem{Dinh:2012bp}
D.~N. Dinh, A.~Ibarra, E.~Molinaro, and S.~T. Petcov, ``{The $\mu - e$
  Conversion in Nuclei, $\mu \to e \gamma, \mu \to 3e$ Decays and TeV Scale
  See-Saw Scenarios of Neutrino Mass Generation},'' {\em JHEP}, vol.~08,
  p.~125, 2012.
\newblock [Erratum: JHEP09,023(2013)].

\bibitem{Dinh:2019jdg}
D.~N. Dinh, D.~T. Huong, N.~T. Duy, N.~T. Nhuan, L.~D. Thien, and P.~Van~Dong,
  ``{Flavor changing in the flipped trinification},'' {\em Phys. Rev. D},
  vol.~99, no.~5, p.~055005, 2019.

\bibitem{Huong:2019vej}
D.~T. Huong, D.~N. Dinh, L.~D. Thien, and P.~Van~Dong, ``{Dark matter and
  flavor changing in the flipped 3-3-1 model},'' {\em JHEP}, vol.~08, p.~051,
  2019.

\bibitem{Chanowitz:1985ug}
M.~S. Chanowitz and M.~Golden, ``{Higgs Boson Triplets With M ($W$) = M ($Z$)
  $\cos \theta \omega$},'' {\em Phys. Lett. B}, vol.~165, pp.~105--108, 1985.

\end{thebibliography}

\end{document}